\documentclass[prl,10pt,twocolumn,amsmath,amssymb,showpacs,floatfix]{revtex4}

\usepackage{graphicx}

\begin{document}
\title{Intercalant and intermolecular phonon assisted superconductivity in K-doped picene}

\author{Michele Casula,$^{1}$ Matteo Calandra,$^{1}$ Gianni Profeta$^{2}$ and Francesco Mauri$^{1}$}

\address{ $^1$ IMPMC, Universit\'e Paris 6, CNRS, 4 Place Jussieu, 7504 Paris, France\\
$^2$ SPIN-CNR - Dip. di Fisica Universit\`a degli Studi di L'Aquila, 67100 L'Aquila, Italy\\ 
and Max-Planck Institute of Microstructure Physics, Weinberg 2, 06120 Halle, Germany}
\date{\today}
\begin{abstract} 
K$_3$-picene is a superconducting molecular crystal with critical
temperature T$_c=7$K or $18$K, depending on preparation conditions. 
Using density functional theory 
we show that electron-phonon interaction accounts for T$_c ~ 3-8$ K.
The average electron-phonon coupling, calculated by 
including the phonon energy scale in the electron-phonon scattering, 
is $\lambda=0.73$  and  $\omega_{\rm  log}=18.0\,$ meV. 
Intercalant and intermolecular phonon-modes contribute substantially ($40\%$)  
to $\lambda$ as also shown by the isotope exponents of potassium
(0.19) and  carbon (0.31). 
The relevance of these modes 
makes superconductivity in  K-doped picene peculiar and different
from that of fullerenes.
\end{abstract}

\maketitle

Most materials made of organic molecules are not metals as the valence bands are completely filled and separated by the conduction band.
However, the discovery in 1980 of superconducting
tetra-methyl-tetra-selenium-fulvalene\cite{TMTSF}, 
demonstrated the possibility to induce superconductivity by intercalation of large organic molecules with inorganic anions, serving as acceptors or donors.
This general rule was confirmed by the discovery of superconductivity  in alkali-doped 
fullerenes\cite{GunnarssonRMP} with
transition temperatures ($T_c$) up to $40$ K in Cs$_3$C$_{60}$~\cite{Palstra95}.

The field was renewed recently  with the discovery of
superconductivity in potassium (K)-intercalated molecular-solids 
based on polycyclic aromatic hydrocarbons, such as picene\cite{Mitsuhashi2010} (T$_c=7$K or T$_c=18$K~\cite{kubozono_private}),
phenantrene\cite{Wang2011} (T$_c= 5$K) and coronene\cite{KubozonoCoronene} (T$_c=15$K).
These systems are very appealing as
they exist in a large variety, and it is possible to
tune their properties in many ways. For instance, the chemistry and functionalization of polycyclic
hydrocarbons is well known and relevant for environmental and
medical issues\cite{HC_book},
the intercalation with alkali atoms is often possible, and
the electrochemical doping in picene field-effect transistors has been
achieved\cite{OkamotoPiceneFet}.

In addition, organic molecular crystals are particularly appealing  for technological applications
 such as  
organic light emitting diodes (OLED), field effect transistors (OFET) or solar cells.
Consequently, the study of  charge injection and carrier transport in these materials are central to the understanding and improvement of such devices. 
In particular, the study of the electron-phonon coupling interaction is crucial to describe the different regimes of charge transport~\cite{Hannewald}.

In molecular crystals, the electrons can couple to three
kinds of vibrations; intramolecular, intermolecular (rigid translations and rotations of molecules) and
intercalant phonons. 
In alkali-doped C$_{60}$, after a long debate,~\cite{alkali_superconductivity,inter_superconductivity,antropov,GunnarssonRMP,CaponeRMP}, it has been established that superconductivity is driven mainly by intramolecular modes.
This conclusion is also supported by the negligible Rb isotope effect
measured in Rb$_3$C$_{60}$\cite{Ebbesen92b,Burk}. However the
situation
could be substantially different in K-doped Picene where a more
prominent role of 
intercalant modes in superconductivity is suggested by the significant hybridization of the 
low-energy bands with K atoms observed in density functional theory
(DFT)\cite{Kosugi09}. Indeed, 
the electronic structure of K-doped picene is different from
that obtained by a rigid-doping of the picene crystal without explicit K ions\cite{Kosugi09}.
Thus, at the moment, it is still unclear (i) if superconductivity in 
K-doped picene is phonon-mediated, (ii) if the intramolecular
phonons alone provide enough coupling or if intercalant and intermolecular
vibrations should be also considered. 

These questions were first addressed in an early DFT calculation~\cite{Subedi} where it was found a sizable
electron-phonon coupling, dominated by high-energy intramolecular in-plane carbon (C) vibrations 
at ~1000 cm$^{-1}$. 
However, this work uses a rigid-band doping approximation of the
pristine picene crystal structure and thus neglects
the molecular relaxation after intercalation, the K contribution to
the electron-phonon coupling, and the screening of the electron-phonon
deformation-potential by the metallic bands. Furthermore  
only modes above 100 cm$^{-1}$ were considered.

In this letter, we perform first-principles DFT calculations  on the
superconducting 
properties of K$_3$-picene. Removing all possible a-priori
approximations, 
and in contrast to~\cite{Subedi},
we demonstrate the critical role played by the intercalant
and intermolecular modes in the electron-phonon coupling mechanism
to account for the experimental $T_c$. 
We give a numerical prediction of the isotope exponents 
in the molecular crystal and relate them to the  phonon-mediated superconductivity.

We compute the electronic structure of K$_3$-picene within 
DFT in the local density approximation (LDA)\cite{footnote1}.
We relax the molecular geometry
in the P$2_1$ symmetry in the experimental unit
cell~\cite{Mitsuhashi2010,footnote2},
obtaining a structure similar to that of \cite{Kosugi09}
with the molecules arranged in a herringbone configuration and the intercalants aligned 
parallel to the molecular planes\cite{footnote3}.

\begin{figure}[htp]
\includegraphics[width=\columnwidth]{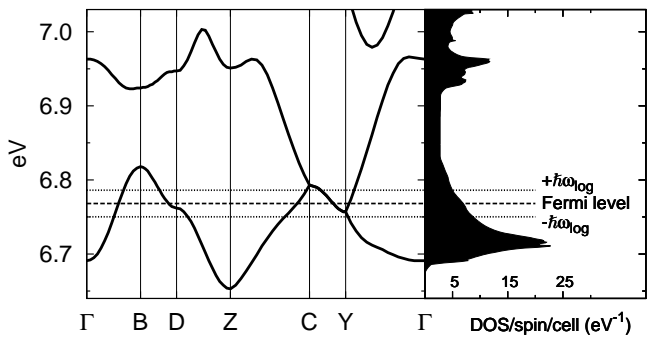}
\caption{
DFT-LDA band structure of K$_3$-picene near the Fermi level. 
The DOS per spin per cell (comprising 2 picene molecules)  is shown in the right panel. Note the strong DOS
variation in a range of $\pm \hbar \omega_{log}$ from the Fermi
energy, with the characteristic phonon frequency $\omega_{log}=18$ meV
} \label{fig:bandsdos} 	
\end{figure}
The band structure of the $K_3$ picene is reported in Fig.~\ref{fig:bandsdos}.
The molecular nature of the compound gives rise to very narrow bands
$\epsilon_{\textbf{k}n}$, with bandwidth $\Delta \epsilon$ of few tenths of eV.
The Fermi level $\epsilon_F$ in the K-doped picene crosses the bands
originating from the LUMO+1 (second Lowest Unoccupied Molecular
Orbital) states of the undoped molecule,
which are entangled to those coming from the LUMO+2 states, and very close to the ones belonging to the LUMO bands.
The bandwidth $\Delta \epsilon$ has the same order of magnitude
as the characteristic phonon frequency $\omega_{ph}$ [the logarithmic phonon average $\omega_{log}$=18 meV, calculated in this work (see below)].
The density of states (DOS) varies substantially
in the range $\left[\epsilon_F-\omega_{log},\epsilon_F+\omega_{log}\right]$, 
as shown in Fig.~\ref{fig:bandsdos}.

\begin{figure}[htp]
\includegraphics[width=\columnwidth]{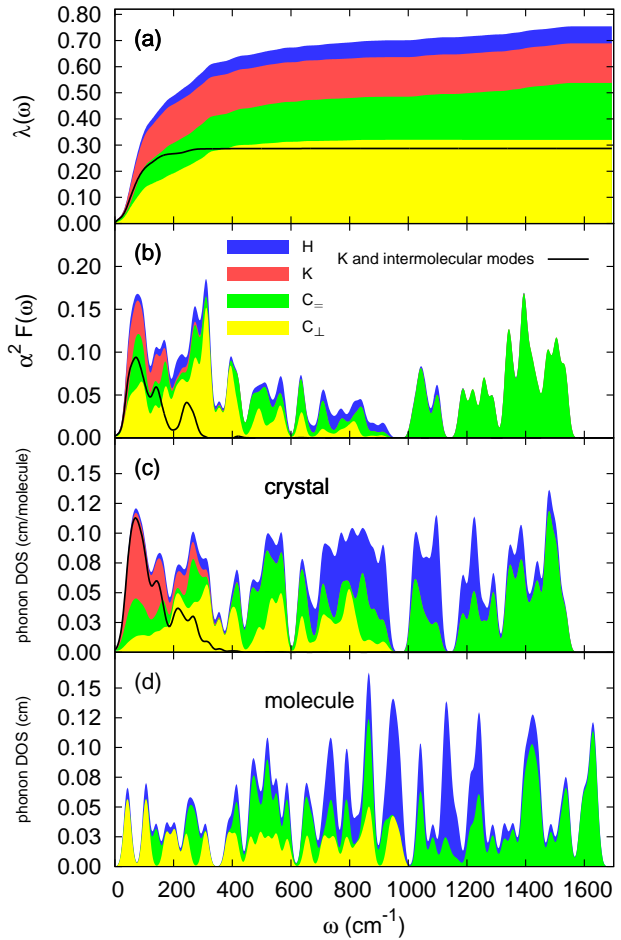}
\caption{
Panels (c) and (d): phonon DOS of the K$_3$-picene crystal  and picene molecule, respectively.
The DOS is projected on phonon subspaces $\textit{S}$ which correspond to hydrogen (H), potassium (K),
out-of-plane and in-plane carbon (C$_\perp$ and C$_=$) modes. The black solid curve is the sum 
of K and intermolecular modes. In the (d) panel the black line is absent as there is no K 
and all phonons are intramolecular, except for the six rototranslational modes
at zero frequency. 
We do not show the C-H stretching modes, which do not couple to
electrons, and are located around 3050 cm$^{-1}$, well above the frequency range plotted.
Panels (a) and (b): $\lambda(\omega)$ and $\alpha^2 F(\omega)$ resolved by projections on H, K, C$_\perp$, and 
 C$_=$ phonons. $\lambda(\omega)$ ramps up by 90\% in the first 300
 cm$^{-1}$.
The black solid line is the sum of K and intermolecular modes, which
contribute to 40\% of the total $\lambda$. 
} \label{fig:lambda_dec} 	
\end{figure}

We carry out phonon calculations in the density functional perturbation theory framework\cite{baroni}.
We consider phonon momenta on a $N_q=2\times2\times2$ grid. We get dynamically stable phonons,
signaling a well converged phonon dispersion $\omega_{\textbf{q}\nu}$ and no structural instabilities in the chosen geometry.
In Fig.\ref{fig:lambda_dec}(c) we report the projected phonon DOS,
$\rho_{\textit{S}}(\epsilon)=\sum_{\textbf{q}\nu} \textbf{e}^*_{\textbf{q}\nu}
\cdot \mathcal{P}_\textit{S} \textbf{e}_{\textbf{q}\nu} \delta(\omega_{\textbf{q}\nu}-\epsilon)$, 
where the phonon eigenstates $\textbf{e}_{\textbf{q}\nu}$ are 3N-dimensional vectors, with N the number of atoms, and the projector 
$\mathcal{P}_\textit{S}$ is a 3N$\times$3N tensor.
We split the full phonon space into potassium (K), hydrogen (H),
in-plane (C$_{=}$) and out-of-plane C$_{\perp}$ carbon $\textit{S}$-subspaces (the plane 
is the one containing each picene molecule). We also consider
partitioning into K, 
intermolecular and intramolecular vibrations. 
The projections reveal  the presence of the K and intermolecular 
modes at low frequencies, while only a very small fraction of intramolecular modes is 
present below 200 cm$^{-1}$. 
On the other hand, the in-plane C and H modes occupy the high-energy part of the DOS.
We compute the phonon spectrum also for the
neutral isolated molecule, 
Fig.\ref{fig:lambda_dec}(d). The low-energy spectrum of molecule and crystal differ
considerably.
In the solid the rigid-body rototranslations (intermolecular modes)
are spread up to 300 cm$^{-1}$, while in the molecule they are pinned at zero frequency. 
In addition we observe in the crystal a significant spectral weight transfer of the 
low-energy out-of-plane C oscillations to higher frequencies.

For each phonon mode $\nu$ with momentum $\textbf{q}$ we compute the electron-phonon interaction:
\begin{eqnarray}
\lambda_{\textbf{q} \nu} & = &  \frac{2}{\omega^2_{\textbf{q} \nu} N(0) N_k} \sum_{\textbf{k},n,m} |g^\nu_{\textbf{k}n,\textbf{k}+\textbf{q}m}|^2 
\nonumber \\
   & \times &  (f_{\textbf{k}n}-f_{\textbf{k}+\textbf{q},m}) ~ \delta(\epsilon_{\textbf{k}+\textbf{q},m}-\epsilon_{\textbf{k}n} - \omega_{\textbf{q}\nu}),
\label{elphon}
\end{eqnarray}
that couples the occupied state $|\textbf{k}, n\rangle$ of momentum
$\textbf{k}$ and band $n$ with the empty state $|\textbf{k}+\textbf{q},
m\rangle$ separated by the phonon energy  $\omega_{\textbf{q}\nu}$. 
The $\textbf{k}$-summation in Eq.~\ref{elphon} is over the Brillouin
zone on a $N_k$ electron-momentum mesh, and $N(0)$ ($=6.2$ eV$^{-1}$) 
is the electron DOS per spin per cell at the Fermi level.
The electron-phonon matrix elements are $g^\nu_{\textbf{k}n,\textbf{k}+\textbf{q}m}=
\sum_s \textbf{e}^s_{\textbf{q}\nu} \cdot
\textbf{d}^s_{nm}(\textbf{k},\textbf{k}+\textbf{q})/\sqrt{2 M_s
  \omega_{\textbf{q}\nu}}$, 
where $\textbf{e}^s_{\textbf{q}\nu}$ is the 3-dimensional $\textbf{q}\nu$ eigenphonon component on
the $s$-th atom with mass $M_s$, and
$ \textbf{d}^s_{nm}(\textbf{k},\textbf{k}+\textbf{q}) =  \langle
\textbf{k}+\textbf{q},m | \delta V/ \delta
u_{\textbf{q}s}|\textbf{k},n\rangle$, with
$ \delta V/ \delta u_{\textbf{q}s}$ the
deformation potential  for a given phonon displacement $\textbf{q}s$
of the $s$-th atom.
In Eq.~\ref{elphon}, $f_{\textbf{k}n}$ are Fermi functions depending on the temperature $T$, and the expression for $\lambda_{\textbf{q}\nu}$ has been evaluated 
by a $T \rightarrow 0$  extrapolation.
In the  ``adiabatic'' limit, namely for $\omega_{ph} \ll
\Delta\epsilon$, where $\Delta \epsilon$ is the bandwith,
the expression for $\lambda_{\textrm{q}\nu}$ in Eq.~\ref{elphon} reduces to the one proposed by Allen\cite{allen}, and generally used in
previous electron-phonon estimates:
\begin{equation}
\lambda^{AD}_{\textbf{q} \nu} =  \frac{2}{\omega_{\textbf{q} \nu} N(0) N_k} \sum_{\textbf{k},n,m} |g^\nu_{\textbf{k}n,\textbf{k}+\textbf{q}m}|^2   
\delta(\epsilon_{\textbf{k},n})  \delta(\epsilon_{\textbf{k}+\textbf{q},m}).
\label{elphon_allen}
\end{equation}
We are going to dub $\lambda^{AD}$ in the Equation above as ``adiabatic'', while $\lambda$ 
in Eq.~\ref{elphon} as ``non-adiabatic''. 

The electron-phonon matrix elements
$g^\nu_{\textbf{k}n,\textbf{k}+\textbf{q}m}$ are computed
on a highly dense electron-momentum grid by means of    
the recently developed Wannier interpolation formalism\cite{marzari,Mostofi,matteo_wannier,giustino_wannier,footnote4}.
This allows one to perform converged calculations with a delta function Gaussian smearing
$\sigma$ much smaller than the bandwidth $\Delta \epsilon$~\cite{footnote5}.
Surprisingly, 
the exact zero-smearing extrapolation leads to a strong
electron and phonon-momentum dependence in $g$, and so in $\lambda$ and $\lambda^{AD}$. 
This is in contrast with the
conventional molecular crystal picture, where $g$ and the
electron-phonon coupling are independent of the electron and phonon momenta.
To quantify this dependence, we evaluated  the  coupling strength
average per molecule,  given
by $2 \lambda^{AD}_{\textbf{q}} N(0) /J_{\textbf{q}}$,  
where $J_{\textbf{q}}= \sum_{\textbf{k},n,m} \delta(\epsilon_{\textbf{k},n})  \delta(\epsilon_{\textbf{k}+\textbf{q},m}) / N_k $
is the nesting factor, $\lambda^{AD}_{\textbf{q}}$ is 
summed over the phonon modes, and the factor of 2 accounts for the
number of molecules in the unit cell.
One obtains the phonon-momentum integrated average-value of 172 meV, with
large variations of $\pm 40 \%$ as a function of the phonon-momentum
~\cite{additional}. 
A detailed analysis of the dependence of $\lambda$ on ${\textbf{q}}$ and $\sigma$ is presented in~\cite{additional}.

\begin{table} 
\begin{tabular}{|l|c|c|c|c|c|c|}
\hline
                   &      $\lambda$   & $\omega_\textrm{log}$(meV) & $T_c$(K)  &   $\alpha$(C)  & $\alpha$(K)  & $\alpha$(H) \\
                   &                  &                     &      $\mu^*\!\in\!\left[0.1,0.2\right]$        &          &          &            \\
\hline
adiabatic           &     0.88        &    25               & 8-16        &          0.32         &     0.13         &    0.05               \\
\hline
non-adiabatic       &     0.73        &    18               & 3-8         &          0.31         &     0.19         &    0.00               \\
\hline
\end{tabular}
\caption{  
$\lambda$, phonon frequency logarithmic average $\omega_{log}$, 
McMillan critical temperature $T_c$ (obtained with $\mu^*$ in the range of 0.1-0.2), and
isotope exponents (for $\mu^*=0.14$), computed based on the electron-phonon coupling expression in Eq.~\ref{elphon} (non-adiabatic), 
which explicitly includes the phonon energy $\omega$ in 
the electron-phonon scattering process, and the one in Eq.~\ref{elphon_allen} (adiabatic), 
which is a $\omega_{ph}/\Delta\epsilon \rightarrow 0$  approximation of Eq.~\ref{elphon}. 
The ``non-adiabatic'' values are our best estimates of the various quantities.
\label{data}
}
\end{table}

In Tab.~\ref{data},  we report our results for the
electron-phonon coupling.
There is a significant difference between the
``adiabatic'' and ``non-adiabatic'' values, as $\omega_{ph} \sim \Delta\epsilon$.
Therefore, one must keep the full expression in Eq.~\ref{elphon} for
an accurate estimate of $\lambda$. 
In Tab.~\ref{data}, we also report the phonon frequency logarithmic average ($\omega_{log}$).

An estimate of $T_c$ is given by the McMillan 
formula~\cite{mcmillan}, which requires $\lambda$, $\omega_{log}$, and the screened Coulomb pseudopotential $\mu^*$.
With $\mu^*\in\left[0.1,0.2\right]$, we obtain $T_c$ in the range $3-8$ K, 
which includes the value of 7 K reported in \cite{Mitsuhashi2010} for one of the two phases of
K$_3$-picene~\cite{kubozono_private}.
Caution must be taken in using the McMillan formula, as 
the on-site molecular correlation can be important with respect to the bandwidth\cite{Giovannetti, Subedi}, 
and also ``non-adiabatic'' effects in the vertex corrections can invalidate the Migdal's theorem\cite{pietronero}.
However, our estimate of $T_c$ is an indication that the electron-phonon coupling is strong enough to explain 
the mechanism behind the manifestation of the superconducting phase.
We computed also the isotope exponents $\alpha(X)=-d \log T_c / d \log M_x$ (see Tab.~\ref{data}) for all constituents of 
the molecular crystal. It turns out that the role of H is negligible, while the large exponent of potassium (K) points
toward the important role played by the intercalant in the electron-phonon coupling to set the value of $T_c$.

For a given value of $\mu^*$ our predicted $T_c$ are more than 8 times
smaller than that predicted with $\lambda$ and $\omega_{log}$ found in~\cite{Subedi}. 
This comes mainly from the value of $\omega_{log}$ (1021 cm$^{-1}$) in~\cite{Subedi}, very
different from our best estimate of 18 meV (145 cm$^{-1}$).
This large discrepancy may come from the 
intercalation driven structural change not considered in that work.
Indeed, the electronic DOS of K$_3$-picene, computed in the rigid-doping approximation with
the undoped geometry, is 14.12 eV$^{-1}$ per spin per
cell~\cite{Subedi,Kosugi09}, a value 2.3 larger than that found here
and in~\cite{Kosugi09} for the relaxed geometry with K.
In addition, \cite{Subedi} neglects
the screening of the electron-phonon interaction by the metallic
bands, the contribution of K phonons and vibrations of energy lower than 100 cm$^{-1}$.

To understand our findings, we compute the Eliashberg function
$\alpha^2 F(\omega) = \sum_{\textbf{q}\nu} \lambda_{\textbf{q}\nu}  \omega_{\textbf{q} \nu} ~ \delta(\omega - \omega_{\textbf{q} \nu}) / (2N_q)$,
plotted in Fig.~\ref{fig:lambda_dec}(b), and the integral 
$\lambda(\omega)= 2 \int_0^\omega d\omega^\prime \alpha^2 F(\omega^\prime)/\omega^\prime$, 
shown in Fig.~\ref{fig:lambda_dec}(a),
which gives the total electron phonon coupling $\lambda=\sum_{\textbf{q}\nu} \lambda_{\textbf{q}\nu} / N_q$
in the $\omega \rightarrow \infty$ limit. 
We note that $\lambda(\omega)$ converges very rapidly to a large
fraction of the total value in the first 300 cm$^{-1}$.
%Neglecting modes below 100 cm$^{-1}$ as done in
%Ref.~\cite{Subedi} with the adiabatic expression 
%leads to $\lambda=0.54$ and $\omega_{log}$=53 meV (427 cm$^{-1}$).

We then decompose $\lambda$ into a set of
phonon subspaces $\textit{S}$, as we did for the phonon DOS, as in
\cite{cac6}.
The evaluation of the projected electron-phonon matrix elements
$g_{\textit{S}} =  \sum_s (\mathcal{P}_\mathit{S} \textbf{e}_{\textbf{q}\nu})^s \cdot
\textbf{d}^s_{nm}(\textbf{k},\textbf{k}+\textbf{q})/\sqrt{2 M_s
  \omega_{\textbf{q}\nu}}$
follows straightforwardly, and the corresponding
projected $\lambda^{\textit{S},\textit{S}^\prime}_{\textbf{q} \nu} $ reads:
\begin{eqnarray}
\lambda^{\textit{S},\textit{S}^\prime}_{\textbf{q} \nu} & = &
\frac{2}{\omega^2_{\textbf{q} \nu} N(0) N_k } 
\sum_{\textbf{k},n,m} g_{\textit{S}} ~ g^\star_{\textit{S}^\prime}
\nonumber \\
   & \times &  (f_{\textbf{k}n}-f_{\textbf{k}+\textbf{q},m}) ~ \delta(\epsilon_{\textbf{k}+\textbf{q},m}-\epsilon_{\textbf{k}n} - \omega_{\textbf{q}\nu}),
\label{elphon_proj}
\end{eqnarray}
where for the sake of readability we dropped out the momentum and orbital indexes from $g_{\textit{S}}$.
$\lambda = \sum_{\textit{S},\textit{S}^\prime}
\lambda^{\textit{S},\textit{S}^\prime}=\sum_{\textit{S},\textit{S}^\prime}\sum_{\textbf{q},\nu} \lambda^{\textit{S},\textit{S}^\prime}_{\textbf{q} \nu} $, and
the contribution of each subspace $\textit{S}$ is computed as  $\sum_{\textit{S}^\prime} \lambda^{\textit{S},\textit{S}^\prime}$,
where we add both the diagonal term and the usually very small
off-diagonal contributions.
The sum over K and intermolecular modes is reported in Fig.~\ref{fig:lambda_dec} as a black solid line. 
It gives more than $40\%$ of the total $\lambda$, with half of it due to the alkali intercalant, and
the corresponding $\alpha^2 F(\omega)$ localized at small frequencies (up to 300 cm$^{-1}$).
The intramolecular phonons contribute to the remaining $60\%$, with the main role played by
the out-of-plane C modes. This is a remarkable result, and shows that 
K-doped picene is far from being a prototype molecular crystal. The
electron-phonon coupling is only partially supported by intramolecular phonons, and the
$\textbf{q}$-dependence of the electron-phonon coupling is strong.

To summarize, we found that in the K$_3$-picene the electron-phonon 
coupling is large enough to explain the experimentally measured $T_c$ of 7 K.
Despite the molecular nature of the crystal, a significant contribution 
to the coupling is given by the dopant and intermolecular phonon modes,
which account for 40\% of the total $\lambda$.
This has a strong impact on the isotope exponent of potassium,
whose value turns out to be large (close to 0.20), and
represents an experimentally-accessible signature of
the importance of non-intramolecular modes.
Our work shows that there are fundamental differences 
between the families of picene and fullerene.
The demonstration of the importance of coupling between the electrons at the Fermi level with  
intermolecular and intercalant low-energy phonon modes will have also a relevant contribution to the 
developing of transport theories, usually accounting for the  local electron-phonon coupling (intramolecular coupling) and considered in a small-polaron theory of narrow bands. 
 Extensions to include non-local couplings (intermolecular coupling) are required\cite{hannewald2}.
 This opens the way to a more comprehensive
view on superconductivity, transport, and other phenomena
in the larger and larger group of recently discovered materials
based on metal-intercalated aromatic hydrocarbons.

We acknowledge Y. Kubozono for useful discussions. Computer allocation
has been supported by CINECA-HPC ISCRA and EU DEISA-SUPERMAG.

%Unused bibitems


\begin{thebibliography}{00}
\bibitem{footnote1}
We use the {\textsc quantum-espresso}\cite{qe}  code. K, C, and H atoms are 
described by ultrasoft pseudopotentials. 
The plane-wave (PW) cutoff is 60 Ry for the
wave-function, and 600 Ry for the charge.  A $4 \times 4 \times 4$
electron-momentum grid and a Methfessel-Paxton smearing  of 0.03 Ry are
used in the electronic integration.
\bibitem{footnote3}
Internal coordinates are reported in the additional materials~\cite{additional}.
\bibitem{footnote4}
Wannierization has been performed on a $4\times4\times4$
electron-momentum mesh, by including the LUMO, LUMO+1, and LUMO+2
 states. Both the long-range hopping and inclusion of the first three LUMO's are needed to get 
localized orbitals and Wannierized bands in a perfect agreement with those computed in the PW (plane waves) basis set
in a window of 0.7 eV around the Fermi level.
\bibitem{footnote5}
We use  
a $120 \times 120 \times 120$ electron-momentum grid, which guarantees
convergence down to the temperature and smeating used in the present calculations ($T=$150K and $\sigma=$4.3 meV, with $T = 3\sigma$).
\end{thebibliography}
\end{document}

\title{Supplemental material for the paper: Intercalant and intermolecular phonon assisted superconductivity in K-doped picene}

\author{Michele Casula,$^{1}$ Matteo Calandra,$^{1}$ Gianni Profeta$^{2}$ and Francesco Mauri$^{1}$}

\address{ $^1$ IMPMC, Universit\'e Paris 6, CNRS, 4 Place Jussieu, 7504 Paris, France\\
$^2$ SPIN-CNR - Dip. di Fisica Universit\`a degli Studi di L'Aquila, 67100 L'Aquila, Italy\\ 
Max-Planck Institute of Microstructure Physics, Weinberg 2, 06120 Halle, Germany}

\maketitle

\section{Geometry} 

We report the coordinates
of the 78  atoms in the monoclinic unit cell with $a,b,c = 8.707,5.912,12.97 {\rm \AA}$, $\beta=92.77^o$.
In the Table, they are split into two groups, one for each
K$_3$-picene  molecule, related each other by the P2$_1$ symmetry operations.
The  atomic symbols are in the first column. The coordinates are in crystal
units.

\begin{table}[hb!]
{\footnotesize
\begin{tabular}{|l|c|c|c||c|c|c|}
\hline
 \multicolumn{1}{|l}{}  &  \multicolumn{3}{|c||}{molecule 1} &  \multicolumn{3}{|c|}{molecule 2} \\
\hline
C & 0.4556221 & 0.6985366 & 0.5225742 & -0.4556221 & 0.1985366 & -0.5225742 \\
C & 0.3666947 & 0.7916786 & 0.5978784 & -0.3666947 & 0.2916786 & -0.5978784 \\
C & 0.4078295 & 0.6965432 & 0.4186143 & -0.4078295 & 0.1965432 & -0.4186143 \\
H & 0.5603062 & 0.6067382 & 0.5464734 & -0.5603062 & 0.1067382 & -0.5464734 \\
C & 0.2680166 & 0.7885757 & 0.3834603 & -0.2680166 & 0.2885757 & -0.3834603 \\
H & 0.4802802 & 0.6105565 & 0.3645861 & -0.4802802 & 0.1105565 & -0.3645861 \\
C & 0.1758863 & 0.9015619 & 0.4566338 & -0.1758863 & 0.4015619 & -0.4566338 \\
C & 0.2192085 & 0.7895931 & 0.2754507 & -0.2192085 & 0.2895931 & -0.2754507 \\
C & 0.0759228 & 0.8943093 & 0.2435217 & -0.0759228 & 0.3943093 & -0.2435217 \\
C & 0.3100630 & 0.7014734 & 0.1996684 & -0.3100630 & 0.2014734 & -0.1996684 \\
C & 0.2617866 & 0.6976639 & 0.0950588 & -0.2617866 & 0.1976639 & -0.0950588 \\
H & 0.4193296 & 0.6208887 & 0.2225564 & -0.4193296 & 0.1208887 & -0.2225564 \\
C & 0.1210633 & 0.7886946 & 0.0646123 & -0.1210633 & 0.2886946 & -0.0646123 \\
H & 0.3366770 & 0.6247067 & 0.0390565 & -0.3366770 & 0.1247067 & -0.0390565 \\
C & 0.0298681 & 0.8877619 & 0.1371182 & -0.0298681 & 0.3877619 & -0.1371182 \\
H & 0.0819831 & 0.7857442 & -0.0168198 & -0.0819831 & 0.2857442 & 0.0168198 \\
H & -0.0771974 & 0.9716713 & 0.1101440 & 0.0771974 & 0.4716713 & -0.1101440 \\
C & -0.0065371 & 1.0089684 & 0.3165922 & 0.0065371 & 0.5089684 & -0.3165922 \\
C & 0.0469952 & 1.0175338 & 0.4200294 & -0.0469952 & 0.5175338 & -0.4200294 \\
H & -0.1098166 & 1.1064455 & 0.2929669 & 0.1098166 & 0.6064455 & -0.2929669 \\
H & -0.0159085 & 1.1243530 & 0.4725799 & 0.0159085 & 0.6243530 & -0.4725799 \\
C & 0.2237032 & 0.8990497 & 0.5655413 & -0.2237032 & 0.3990497 & -0.5655413 \\
C & 0.1396179 & 1.0059855 & 0.6385974 & -0.1396179 & 0.5059855 & -0.6385974 \\
C & 0.1916193 & 1.0169922 & 0.7421052 & -0.1916193 & 0.5169922 & -0.7421052 \\
H & 0.0315715 & 1.0922925 & 0.6160368 & -0.0315715 & 0.5922925 & -0.6160368 \\
C & 0.3247277 & 0.9053904 & 0.7780469 & -0.3247277 & 0.4053905 & -0.7780469 \\
H & 0.1258220 & 1.1133177 & 0.7972359 & -0.1258220 & 0.6133177 & -0.7972359 \\
C & 0.4177289 & 0.7943654 & 0.7053137 & -0.4177289 & 0.2943654 & -0.7053137 \\
C & 0.3729009 & 0.9021317 & 0.8836312 & -0.3729009 & 0.4021317 & -0.8836312 \\
C & 0.5070607 & 0.7918688 & 0.9174399 & -0.5070607 & 0.2918688 & -0.9174399 \\
H & 0.2970889 & 0.9733633 & 0.9403859 & -0.2970889 & 0.4733633 & -0.9403859 \\
C & 0.6002138 & 0.6930755 & 0.8478521 & -0.6002138 & 0.1930755 & -0.8478521 \\
H & 0.5342170 & 0.7751262 & 1.0001134 & -0.5342170 & 0.2751262 & -1.0001134 \\
C & 0.5569571 & 0.6993506 & 0.7422534 & -0.5569571 & 0.1993506 & -0.7422534 \\
H & 0.7001736 & 0.5946126 & 0.8756002 & -0.7001736 & 0.0946126 & -0.8756002 \\
H & 0.6292360 & 0.6123495 & 0.6882365 & -0.6292360 & 0.1123495 & -0.6882365 \\
K & 0.0968372 & 0.3605062 & 0.1955914 & -0.0968372 & -0.1394938 & -0.1955914 \\
K & 0.3513565 & 0.3915392 & 0.8277846 & -0.3513565 & -0.1084608 & -0.8277846 \\
K & 0.2382784 & 0.3669465 & 0.5169973 & -0.2382784 & -0.1330535 & -0.5169973 \\
\hline
\end{tabular}
}
\end{table}

\section{Broadening and $\mathbf{q}$-dependence of $\lambda$}

In the Figure, we present  the 
phonon momentum $\textbf{q}$-resolved $\lambda$'s, evaluated as in 
Eq. 2 in the paper, as a function of the Gaussian smearing
$\sigma$ of the delta functions. The plotted
results are obtained by using a $120^3$ electron $\textbf{k}$-point
mesh. The calculations performed with $60^3$ $\textbf{k}$-point mesh
give identical results down to $T=$150K and $\sigma=$4.3 meV, with $T = 3\sigma$.
The $\textbf{q}$ vectors are reported in crystal units. 
The exact zero-smearing extrapolation leads to a strong
$\textbf{q}$-dependence in $\lambda$, which can be seen 
only for $\sigma$ much smaller than the bandwidth $\Delta \epsilon$ ($ \ll 0.1$ eV).
In the inset, we show the $\textbf{q}$-integrated ``adiabatic'' and
``non-adiabatic'' $\lambda$. The ``non-adiabatic'' $\lambda$
is extrapolated with respect to both $\sigma$ \emph{and}
$T$.

Note the divergence of $\lambda^{AD}_\textbf{q}$ at the zone center. This occurs
in general in presence of optical modes, because of the divergence of
the nesting factor $J_\textbf{q}$ at $\textbf{q}$=0. For this reason,
in the ``adiabatic'' $\textbf{q}$-averaged $\lambda^{AD}$ the $\Gamma$
point has been excluded. In the ``non-adiabatic'' case,
$\lambda_\textbf{q}$ is well behaved also at $\Gamma$.

\begin{figure}[htp]
\includegraphics[width=\columnwidth]{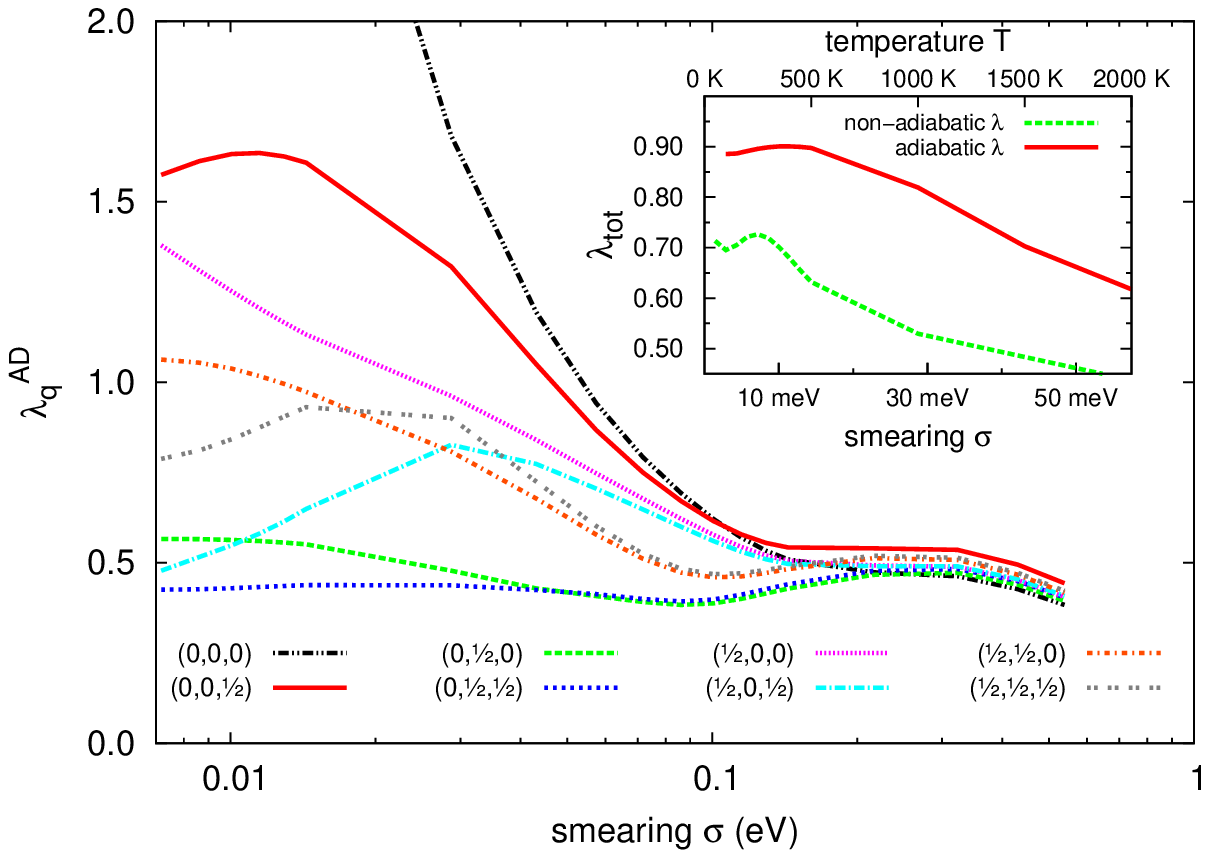}
\end{figure}

\section{Coupling strength average}

From $\lambda^{AD}_\mathbf{q}$, we can quantify the  $\textbf{q}$-dependent coupling strength
average per molecule, by the formula $2 \lambda^{AD}_{\textbf{q}} N(0) /J_{\textbf{q}}$,  
where $J_{\textbf{q}}= \sum_{\textbf{k},n,m} \delta(\epsilon_{\textbf{k},n})  \delta(\epsilon_{\textbf{k}+\textbf{q},m}) / N_k $
is the nesting factor, $\lambda^{AD}_{\textbf{q}}$ is 
summed over the phonon modes, and the factor of 2 accounts for the
number of molecules in the unit cell.
One obtains the values (in meV) $\{$ 183, 193, 131, 133, 239, 146, 191,
  161 $\}$ for the $\textbf{q}$-vectors (in crystal units) $\{$
  $(0,0,0)$,
$(0,0,\frac{1}{2})$, 
$(0,\frac{1}{2},0)$, 
$(0,\frac{1}{2},\frac{1}{2})$, 
$(\frac{1}{2},0,0)$, 
$(\frac{1}{2},0,\frac{1}{2})$, 
$(\frac{1}{2},\frac{1}{2},0)$, 
$(\frac{1}{2},\frac{1}{2},\frac{1}{2})$ $\}$, respectively.
Note that even at $\Gamma$ the ratio is well defined, as the
divergences of both $\lambda^{AD}_\Gamma$ and $J_\Gamma$ cancel out.